\documentstyle[12pt]{article}
\begin{document}
\hoffset = -1truecm
\voffset = -3truecm
\newcommand{\marginlabel}[1]
{\mbox{}\marginpar{\raggedright\hspace{0pt}#1}}


\title{About Heisenberg Uncertainty Relation}

\author{By E. Schr\"odinger\\
Proceedings of The Prussian Academy of Sciences\\
Physics-Mathematical Section. 1930. XIX, pp.296-303 $^{+}$}
\date{}
\maketitle
\begin{abstract}
The original Schr\"odinger's paper is translated and
annotated in honour of the 70$^{-th}$ anniversary of his
Uncertainty Relation 
[published also in:  Bulg. Journal of  Physics, vol.26, nos.5/6 (1999) pp.193-203]. 
In the annotation it is shown that the Uncertainty Relation
can be written in a complete compact {\it canonical} form.
\end{abstract}

{
\hsize = 16 cm
\parindent = 0 pt
\leftskip = 1 cm
\footnotesize \it ANNOTATION \,\, by A. Angelow $^{++}$,
M.-C. Batoni $^{+++}$\\

The main reason to publish the original Schr\"odinger's paper
in English, is the fact that {\bf no one} of the books on Quantum
Mechanics cites it (see for example [1$^\dagger$-15$^\dagger$]).
Actually, the Schr\"odinger's paper is chiefly based on the notes
of the seminars of
Physics-Mathematical Section of The Prussian Academy, where many
famous physicists worked to establish the underlying basis of
Quantum Theory. Being a kind of internal report, this work remained,
for many years at a certain marginal distance from the physicist
scientific awareness.
Another argument in favour of its oblivion concerns the
enthusiastic discussions, mostly about the physical interpretation of the
uncertainty principle, rather than its mathematical straightforward
derivation. After Schr\"odinger, the very first appearance of
the new uncertainty relation
occurs in the book of Merzbacher. However, he has not pay any attention
to the new term (-- the covariance) and directly derives the Heisenberg
inequality [11$^\dagger$]. The same  embarrassment appeared
in [12$^\dagger$] and [13$^\dagger$].\\
Fortunately,  over the last years the scenario changed:
for example, in the field of Quantum Optics,
which was avocated to demonstrate the  fundamental limit
of Quantum Theory with understanding underlying on that new
term, two monographs [14$^\dagger$,15$^\dagger$] were published,
but the authors missed to cite the original Schr\"odinger's paper.\\
Schr\"odinger's work, originally written in German, was translated only in
russian by A. Rogali [16$^\dagger$], in 1976. 
We would like to emphasize that Schr\"odinger paid special attention
to the new term (covariance) in (9), discussed in details in the second 
paragraph after equation (11). This covariance
allows introducing new classes of states, different from coherent 
and squeezed, as it was done in [20$^{\dagger}$-22$^{\dagger}$],
where proposal for their experimental realization was given.\\
Finally, we would like to demonstrate the essential contribution of
Schr\"odinger's  inequality presenting it in a new compact
form, using a modern terminology from mathematics, rather
than that used in 1930. Let us take the three independent
second-order central moments of the joint quantum distribution
of two variables $A$ and $B$, which are of special interest and
warrant a special notation (see, for ex. [17$^\dagger$,18$^{\dagger}$]):
$$
E[ (A-E[ A] )^2 ] \equiv \overline{(A-\overline{A})^2}\equiv
Var[ A ]\equiv \Delta(A)\,^2   \eqno(1^\dagger)
$$ $$
E[ (B-E[ B])^2]\equiv \overline{(B-\overline{B})^2}\equiv
Var[ B ]\equiv \Delta(B)\,^2 \eqno(2^\dagger)
$$ $$
E\left[
{{(A-E[A])(B-E[B])+(B-E[B])(A-E[A])}\over2}\right]
\equiv Cov[A,B]\equiv \Delta(A,B) \eqno(3^\dagger)
$$
where $E[.]$ means the expectation value.
Obviously, the fourth second-order moment is $Cov[B,A]=Cov[A,B]$,
(respectively $\Delta(B,A)=\Delta(A,B)$).
Note that when the observables $X$ and $Y$ don't commute, the
correct expression for their product is not $XY$, but the symmetrized
one $XY+YX\over2$. We conclude that the covariance is equivalent
to the new term in the Schr\"odinger's inequality:
$$
Cov[A,B]={\overline{AB+BA}\over2}-\overline{A} \, \overline{B}
\eqno(4^\dagger)
$$
Let us construct the so-called [19$^\dagger$] {\bf covariance matrix}
(keeping in mind the non-commuta\-ti\-vi\-ty of the observables in contrast
to [19$^\dagger$]):
$$
{\sigma[A,B]}\equiv\left ( \matrix{
Var[A]                       & Cov[A,B] \cr
Cov[B,A]  & Var[B]
} \right )\quad \equiv
\quad  \left( \matrix{
\Delta(A)\,^2                       & \Delta(A,B) \cr
\Delta(B,A)  & \Delta(B)\,^2
} \right )                     \eqno(5^\dagger)
$$
Now, we can write the Schr\"odinger Uncertainty Principle
( see below-eq.(9) ) in the {\bf canonical} form
$$ det(\sigma[A,B]) \geq {1\over4}{| {\overline{[A,B]}} \,|}^2,
\quad {\it{or,\, for\, position\, and\, momentum}}\quad
det(\sigma[q,p]) \geq {\hbar^2\over4}, \eqno(6^\dagger)$$
and it is easy to see that the uncertainty relation is invariant under
the rotation transformation in the phase space, while the Heisenberg one
is not. We would like to emphasize also that the new term in the inequality
also plays an important role in the method of linear invariants
in Quantum Mechanics, where the covariance is expressed in
terms of the solution of the equation of a non-stationary
two-dimensional harmonic oscilator [20$^{\dagger}$].\\
$\quad$ Translated and annotated in honour of 70$^{-th}$
anniversary of Schr\"odinger Uncertainty Relation.\\
Sofia,\,\,\,\,\, January 1999, February 2008\\
S\~ao Paulo, March 1999.\\
$$
$$}
\vspace{.5cm}
\,\,
\\
\S 1. Recently E. U. Condon and H. P. Robertson \cite{condon}
took into consideration
the generalization of the fundamental principle of the quantum mechanics
- that of the uncertainty - over an arbitrary canonical non-conjugate
couple of physical variables. Trying to reach the same, I arrived at a
slightly wider generalization than the Robertson's one, which is,
in fact stronger than the original Heisenberg inequality.\\

First of all, let us set out what is well known. The state-of-the-art of
the ``interpretation question" is the following: the test domain is a single
specific physical system. The base for the system knowledge that we
dispose of - the catalogue of all that we can assert about the system -
is equivalent to a complex function $\Psi$ in the coordinate space of the
system (it changes in a regular manner in time, but is not important
at the moment). The mathematical correlate of a ``physical variable",
i.e. of a very specific measurement that one might apply to the system, is
a very specific linear Hermitean operator that from each $\Psi$-function
produces an other such a $\Psi$-function. One can calculate the
{\bf expectation value} of the respective measurement from the measure
operator, say $A$, and the given $\Psi$-function:
\begin{equation}\label{eq:1}
\overline A = \int \Psi^* A \Psi dx
\end{equation}
($\Psi^*$ is the complex conjugate, the integration goes over the whole
coordinate space; given that $\Psi$ is constantly normalized, i. e.
$\int\Psi^* \Psi dx = 1$).\\

The meaning of the expectation value is: a mean value by an
unlimited number of measurements, while one must be sure that the
system state is the same before each measurement, not changed by the
measurement itself. In general, all possible statements one can make
about the system are encoded in the {\it expectation values}.
Moreover, one should keep in mind that it is
up to us to choose the marking of reference scale of our measurement
instrument. We can, for example, set a value one only to one scale
division and zero to all other.
There is a specific operator attached to this ``measurement" --
one could name it as an operator in blinkers$^{\dagger\dagger}$,
\,V.Neumann named it identity operator. The respective expectation
is obviously nothing else but the {\bf probability} of the
corresponding measurement value or measurement value intervals. The
$\Psi$-function determines also the total measurement {\bf statistics}.\\

The average error or the mean uncertainty of the value, which belongs
to the operator $A$, is defined as
\begin{equation}\label{eq:2}
\Delta A = \sqrt{\overline{(A - \overline{A})^2}} =
\sqrt{\overline{A^2} - (\overline{A})^2}
\end{equation}
(where in the first of the two expressions $\overline{A}$
should be more precise: $\overline{A}$ multiplied by the
{\bf identity} operator.) It may be proven, that this definition is
not only formally constructed according to the theory of errors, but
$\Delta A$ is {\bf really} the average error of the variable $A$,
when the statistics is defined in the above given way.\\
To prove now, that the product of the uncertainties of
two random variables $A$ and $B$ satisfies the Heisenberg or even
more precise inequality, we need to denote
the following mathematical statements:

        1. the Hermitean character of $A$ implies that the expectation
value (1) is constantly real;

        2. for each Hermitean operator it holds
\begin{equation}\label{eq:3}
\int f A g dx = \int g A^* f dx ,
\end{equation}
i.e., it could be rolled over on the other factor in such
an integral, in this case the operator transforms into its
conjugate form \cite{a_star};

        3. the product of two Hermitean operators is in general not
Hermitean, but it could be split into a ``symmetrical product" and
half of its commutator:
\begin{equation}\label{eq:4}
AB= {{AB+BA}\over2}+{AB-BA\over2}
\end{equation}
The first term is Hermitean, the last one is ``skew Hermitean",
i.e. it becomes Hermitean multiplied by $i=\sqrt{-1}$. The splitting in
many aspects corresponds to the splitting of a random (complex)$^{\dagger}$
number into real and imaginary parts. Immediately from here one might
extract the splitting of the {\bf expectation value} into real and
imaginary parts. The ``expectation value" of every commutator is pure
imaginary.

        4. Finally, we need the so-called Schwartz inequality \cite{weyl}
\begin{equation}\label{eq:5}
(a_1a_1^*+a_2a_2^*+...+a_na_n^*)(b_1b_1^*+b_2b_2^*+...+b_nb_n^*)
\geq |a_1b_1+a_2b_2+...+a_nb_n|^2 ,
\end{equation}
that we shall apply in a limiting case on the continuous range of
the values of both functions $f$ and $g$ in the coordinate space:
$$
\hspace{5cm} \int ff^*dx \cdot \int gg^*dx
\geq \left|\int fg dx\right|^2 . \hspace{4.5cm} (5')
$$
We assume here that specially
\begin{equation}\label{eq:6}
f=B\Psi \qquad g=A^*\Psi^*,
\end{equation}
where $A$ and $B$ are some Hermitean operators and $\Psi$ is
an arbitrary wave function, i.e. an arbitrary continuous and normalized
function in the coordinate space. Using the equation (3) one obtains
\begin{equation}\label{eq:7}
\int \Psi^* B^2 \Psi dx \cdot \int \Psi^* A^2 \Psi dx
\geq \left |\int \Psi^* AB \Psi dx\right |^2 ,
\end{equation}
i.e., in terms of the notation (1)
$$
\hspace{6cm} \overline{A^2} \cdot \overline{B^2}
\geq\left|\overline{AB}\right|^2 . \hspace{6.3cm} (7')
$$
If we decompose the right hand side according to (4), then we get
\begin{equation}\label{eq:8}
\overline{A^2} \cdot \overline{B^2}\geq
\left(\overline{AB+BA}\over2\right)^2+
\left|\overline{AB-BA}\over2\right|^2 .
\end{equation}
This is already the inequality that we need to proof, but only in the
{\bf special case} when $\overline{A}$ and $\overline{B}$ vanish.
In order to arrive at the general case, one should apply (8) and instead
of the operators $A$ and $B$ rather use the following
$$
A-\alpha{\bf 1} \quad {\rm {and}} \quad B-\beta{\bf 1}.
$$
First of all $\alpha$ and $\beta$ must be arbitrary real constants,
\,$\alpha{\bf 1}$ is the (identity)$^{\dagger}$ operator
multiplied by $\alpha$. The resulting inequality is therefore valid:\,\,
1. for an arbitrary $\Psi$, \,\,
2. for every real pair of constants $\alpha,\beta$.
Therefore, there is no limitation on the $\Psi$-function to
influence the choice of the pair of constants and especially to set
$$
\alpha = \overline{A}, \quad \beta = \overline{B}.
$$
Finally, we end up with:
\begin{equation}\label{eq:9}
(\Delta A)^2 (\Delta B)^2\geq
\left({\overline{AB+BA}\over2}-\overline{A} \, \overline{B}\right)^2+
\left|\overline{AB-BA}\over2\right|^2 .
\end{equation}
This is the final form. The first from the two addends on the right hand
side is a new one (to the best of my knowledge). (Without that term
the inequality stands as the one of
H. P. Robertson.) So, the inequality links together {\bf three} quantities:
\, 1. the product of the mean deviations squared,
\, 2. the absolute value squared of half of the mean value
of the commutator, \, 3. a quantity which could be
defined as a square of the mean deviations-product
(the covariance)$^{\dagger}$ in the condition that
non-commutability is taken into account, i.e. the mean
deviations-product must be define as the arithmetic mean of
\begin{equation}\label{eq:10}
\overline{(A - \overline{A})(B - \overline{B})} \quad {\rm and}
\quad \overline{(B - \overline{B})(A - \overline{A})}
\end{equation}
which are the {``mixed"} expressions
{( {$\equiv$} covariances,
see eq.(3{$^{\dagger}$}) )$^{\dagger}$},  completely analogous
to $(\Delta A)^2$ and $(\Delta B)^2$ \, {$^{\dagger\dagger\dagger}$}.\\

One is led to the Heisenberg inequality when the last mentioned quantities
are stricken out in order to make stronger the inequality and $A$, $B$ are
chosen to be canonically conjugate:
$$
AB-BA={h\over2\pi i} .
$$
Then it results in
\begin{equation}\label{eq:11}
{\Delta A}\cdot{\Delta B}\geq{h\over4\pi} .
\end{equation}
On the other hand, it is known that the Heisenberg limit is
{\bf not really} too low, but for some special $\Psi$-functions
{\bf achieves} even higher value \cite{neumann}.
This implies that at least for these special
$\Psi$-functions the (central)$^{\dagger}$ mean deviations-product of
the canonical conjugate operators vanishes. This will be used in \S 2.\\

In the classical theory of errors or fluctuation theory it is well known
that the vanishing of the mean deviations-product  is a {\bf necessary}
(but not sufficient) condition for two values to fluctuate totally
independent one from another. While canonically conjugate quantum
variables have some ``independence" that could mean that some precise
knowledge about one excludes such a knowledge about the other, so one
could perhaps suppose that their mean deviations-product, i. e.
for each $\Psi$-function, has vanishing expectation value.
But this is not the case. Let us consider the two
canonically conjugate operators
$$
A=x \qquad B={h\over2\pi i}{\partial\over\partial x} ,
$$
so we get \cite{infinity}
$$
{{2\pi i}\over h}\,\, {\overline{AB+BA}\over2}=
{1\over2}\int \Psi^*\left[x{\partial\Psi\over\partial x}+
{\partial\over\partial x}\left(x\Psi\right)\right]dx=
{1\over2}\int x\left(\Psi^*{\partial\Psi\over\partial x}-
\Psi{\partial\Psi^*\over\partial x}\right)dx
$$ $$
={1\over2}\int x \Psi^*\Psi{\partial\over\partial x}
\left(ln{\Psi\over\Psi^*}\right)dx \hspace{3.5cm}
$$ $$
\overline A = \int x\Psi^*\Psi dx \hspace{6.5cm}
$$ $$
{{2\pi i}\over h}\,\, \overline{B}=
\int \Psi^*{\partial\Psi\over\partial x}dx=
{1\over2}\int \left(\Psi^*{\partial\Psi\over\partial x}-
\Psi{\partial\Psi^*\over\partial x}\right)dx \hspace{1.9cm}
$$ $$
={1\over2}\int \Psi \Psi^*{\partial\over\partial x}
\left(ln{\Psi\over\Psi^*}\right)dx .\hspace{3.3cm}
$$
Let now $\Psi = r e^{i\Phi}$ with real $\Phi$ and real, non-negative $r$,
which must satisfy the normalizing condition
$$
\int{r^2}dx=1
$$
Then we get:
$$
{{2\pi}\over h}
\left({\overline{AB+BA}\over2}-\overline{A} \, \overline{B}\right)=
\int x r^2{\partial \Phi\over\partial x}dx-
\int x r^2dx \cdot \int r^2{\partial \Phi\over\partial x}dx .
$$
As ${\partial \Phi\over\partial x}$ is any real function and $r^2$
is an absolute non-negative function (not taking into account the
normalizing condition), so we get that in general the right
hand side does not vanish. One needs
for example to choose $r^2$ to be {\bf even} and
${\partial \Phi\over\partial x}$ to be {\bf odd}
(and not identically vanishing), so the deviation product is surely
positive.\\

As is known, the canonically conjugate quantum variable
is not unambiguously defined.
If $B$ is conjugate to $A$ this implies that $B+\varepsilon A$ is as well
($\varepsilon$ is any real number).
With this change the mean deviations-product
changes too, and becomes, as one could easily calculate,
$\varepsilon(\Delta A)^2$. In the same manner, the result will be
$\varepsilon(\Delta B)^2$, if $A$ was changed to
$A+\varepsilon B$. This can always make the deviations-product
equal to zero by changing one of the operators, without
changing their canonical relation. The change depends on the
above shown special $\Psi$-function of course. One can not reach an
identical vanishing of the deviations-product
in such a manner.\\

\S 2. To the discovery of the complete inequality (9) we are led,
by a chance, to the following question, which is interesting by itself.
Let us consider a force free mass point, mass $m$, coordinate $q$,
momentum $p$, Hamilton-function $H={p^2\over2m}$.
I must undertake simultaneous measurements of the coordinate and
momentum at {\it ``time zero"}, with highest possible precision,
i.e. so that
\begin{equation}\label{eq:12}
{\Delta q_0} \,\, {\Delta p_0} ={h\over4\pi} .
\end{equation}
Further I could distribute the error on $q_0$ and $p_0$ {\bf so}
that for a {\bf given} later time point $t$, could  achieve
the most precise {\bf place}.
This means $\Delta q$ to become the least possible. We use for this
purpose the very convenient ``$q$-number-method", which is in a
methodical manner opposing to the wave mechanics. I would like to
elucidate shortly on it here, repeting what is well known.
For the theorist working on a wave mechanics the operator, which
corresponds to a specific physical variable, does not change in time.
If one wants to know the mathematical expectation value for
this variable, one calculates the $\Psi$-function
for this later moment from the ``time-dependent wave equation".
Then one applies the corresponding operator,
which mentioned above, which is the same for every moment.
On the other side the $q$-number-theorist has to operate with
one single $\Psi$-function at one single chosen moment, once and for all.
However it is unnecessary to express any statement for it, once the moment
is totally arbitrary chosen. One assumes, that the {\bf operators}
are time dependent and we may ask instead: how does the operator
change itself in time, i.e. which operator should be applied on
the {\bf original} $\Psi$-function, in order to calculate the
mathematical expectation of the respective value at the time $t$?\\

Here we point out, that one may calculate the operators
(or $q$-numbers, or matrices) almost as the usual numbers,
and indeed, their change in time is determined by the
{\bf equation of motion of the classical mechanics}.
The only difference is that, occasionally, when it is the case,
one should pay special attention to an eventual non-commutability of the
operator multiplication.\\

So, in this present simple case, the integration of the
equation of motion reads:
$$
q=q_0+{t\over m}p_0 .
$$
One can directly make from it the mean square deviation of the
coordinate, $(\Delta q)^2$, for every moment $t$:
$$
(\Delta q)^2=\overline{\left(q_0+{t\over m}p_0\right)^2}-
\left(\,\, \overline{q_0+{t\over m}p_0}\,\, \right)^2 $$ $$
\hspace{4cm} =(\Delta q_0)^2+{2t\over m}
\left(\,\, \overline{q_0p_0+p_0q_0\over2}-\overline{q_0}
\, \overline{p_0}\,\, \right)^2+{t^2\over m^2}(\Delta p_0)^2 .
$$
The middle term above is essentially the mean deviations-product of
$q_0$ and $p_0$, which vanishes, in accordance with the
prediction, when $q_0$ and $p_0$ are determined with optimal precision.
Then we simply have
$$
(\Delta q)^2=(\Delta q_0)^2+{t^2\over m^2}(\Delta p_0)^2
$$
or using (12)
$$
(\Delta q)^2=(\Delta q_0)^2+
\left(h\over4\pi\right)^2{t^2\over m^2}{1\over(\Delta q_0)^2},
\qquad {\rm see \,\, Ref.\cite{delta_q}} .
$$
This expression becomes a minimum for that value of $(\Delta q_0)^2$,
which makes both addends on the right hand side equal, i.e. for
$$
(\Delta q_0)^2={ht\over4\pi m} ;
$$
$(\Delta q)^2$ is then exactly twice the value of $(\Delta q_0)^2$, i. e.
\begin{equation}\label{eq:13}
\Delta q=\sqrt{ht\over2\pi m} .
\end{equation}
It seems to me, that this {\bf final result} is likely to have two
points of interest. First, the proportional relation with
the {\bf square root} from the time,
which makes allusion to well known classical deviation principles.
Secondly, that the statement has an remarkable absolute character,
namely, the precision attainable in a later moment depends only on
intermediate time and not on the initial momentum. For a free electron,
as an example, one might give a place prognosis for the end of
the first second on the bases of {\bf already} taken measurements of
position and momentum, in the most favorable case with a precision of
$1cm$, quite independent of whether the electron is fast or
slow \cite{slow}.\\

Of course, at a very high speed this will be changed as it
should take into consideration the relativity theory.
I believe that this could occur by the following simple
considerations. The equation (13) is applied to
the rest reference system of a point mass. Let $m$  be the
rest mass, $t_r$, will be the internal time, so
\begin{equation}\label{eq:14}
(\Delta q)_r=\sqrt{ht_r\over2\pi m_r} .
\end{equation}
This is the precision that is attainable for a moving observer when
calculating the position of the point mass of the co-moving system for the
moment, called ``$t_r$ seconds later". When the observer shows his
knowledge through signs in the space, to the ``rest observer"
those signs seem to be nearer to each other in relation
of \, $\sqrt{1-{\beta}^2} : 1$ \, ;
further he ought say looking from his point of view that
the prognosis were made for the time interval
\begin{equation}\label{eq:15}
t={t_r\over\sqrt{1-\beta^2}} ,
\end{equation}
because for him the clock, with which all the statements of the moving
observer were made, seemed run slower than his own clock.
From his point of view, the mean error decreases
\begin{equation}\label{eq:16}
\Delta q=\sqrt{1-\beta^2}\sqrt{ht\sqrt{1-\beta^2}\over2\pi m_r}=
\sqrt{1-\beta^2}\sqrt{ht\over2\pi m} .
\end{equation}
It becomes smaller and comes nearer to zero when the velocity approaches
are nearing the speed of the light.
This happens not only when the mass $m$ goes
to infinity, but also for a series of point masses moving
with an ever growing speed and an ever smaller rest mass
such that all the moving masses $m$ keep the same value $m$.
Even in this case the maximum precision grows unlimited with
the velocity approaching the speed of the light. This is indeed satisfying,
since this is a boundary process that gives a hope to obtain an accurate
statement for a {\bf light quantum}. And this is really true for light
quantum because the Maxwell waves exhibit no dispersion, and preserve
indefinitely long the place precision which they got in the beginning,
and it could indefinitely grow, since the strong momentum dispersion
which is connected with it, does not have bad influence.\\
\\
Reported on the 5th June 1930\\
Joint-Staff Meeting on the 19th June 1930\\
Distributed on the 16th July 1930\\

{-----------------------}\\
$^{\dagger}  )$ -- Note added from the translators.\\
$^{\dagger\dagger})$ -- By this Schr\"odinger means that the operator
does not change the direction and modulus of the vector, as the
horse in blinkers does not change the direction and the speed until
this is not required by the driver.\\
$^{\dagger\dagger\dagger})$ -- Indeed, if one put $B=A$ in
(3$^{\dagger}$), then $Cov[A,A]= Var[A]\equiv {\Delta(A)^2}$.\\
\,
\\
$^{+})$ Reprinted in:
ERWIN SCHR\"ODINGER, Gesammelte Abhandlungen, Band 3, Wien,
Verlag der \"Osterreichischen Akademie der Wissenschaften,
pp.348-356 (1984)\\
$^{++})$ Permanent address: \\
Institute of Solid State Physics, Bulgarian Academy of Sciences,
 72 Trackia Blvd., 1784 Sofia, {\bf Bulgaria}.\\
$^{+++})$ Permanent address: \\
Instituto de Fisica Teorica, Universidade Estadual Paulista,
Rua Pamplona, 145, CEP 01405-900 Bela Vista,
S\~ao Paulo, {\bf Brazil}.\\
\section*{{References} {\rm\normalsize added from translators.}}
\ \
\,
[1$^{\dagger}$] W.Heisenberg,
The physical principles of the quantum
theory, New York, Dover, (1930);\\ \indent
[2$^{\dagger}$] J. Von Neumann,
Mathematical foundations of quantum
mechanics, Princeton, NJ, University Press, (1955);\\ \indent
[3$^{\dagger}$] P.A.M. Dirac, The principles of quantum mechanics. 4th rev.
ed., Oxford, Clarendon Press (1958);\\ \indent
[4$^{\dagger}$] L.Landau, E.Lifshits,
Quantum Mechanics - non-relativistic theory.
3rd ed., Oxford, Pergamon Press, (1977);\\ \indent
[5$^{\dagger}$] L.de Broglie, Heisenberg Uncertainties and
Probabilistic Interpretation of Wave Mechanics,
Dordrecht, Kluwer Academic Publishers (1995) Chapter 8, eqs.(3)
and (8);\\ \indent
[6$^{\dagger}$] W.Price, S.Chissick, W.Heisenberg,
The uncertainty principle and foundations of quantum mechanics:
a fifty years' survey, New York, Wiley, (1977);\\ \indent
[7$^{\dagger}$] A.Bohm, M.Loewe, Quantum mechanics:
foundations and applications. 2nd rev. and enl.ed.,
 New York, NY, Springer-Verlag (1986);\\ \indent
[8$^{\dagger}$] C.Cohen-Tannoudji, B.Diu, F.Laloe, Mecanique quantique,
v.1, Paris, Hermann (1973) ;\\ \indent
[9$^{\dagger}$] L.Schiff, Quantum Mechanics, 3rd ed.,
International series in pure and applied physics,
New York, McGraw-Hill (1968);\\ \indent
[10$^{\dagger}$] A. Messiah, Quantum Mechanics, v.1, North-Holland
Publishing Company, Amsterdam (1961) p.300, eq.(VIII.9);
actually the author had been very close to the general relation, only
that he had not taken the centralized covariance, see the proof below
eq.(VIII.9);\\ \indent
[11$^{\dagger}$] E.Merzbacher, Quantum Mechanics, 3rd ed.,New York, NY,
Wiley (1998) p.219, eq. (10.58); the same situation with the 1st and 2nd
editions;\\ \indent
[12$^{\dagger}$] A.Das, C.Melissinos, Quantum mechanics: a modern
introduction, New York, NY., Gordon \& Breach, (1986);\\ \indent
[13$^{\dagger}$] J.Sakurai, S.Tuan, Modern quantum mechanics. Rev. ed.,
Reading, MA, Addison - Wesley, (1994);\\ \indent
[14$^{\dagger}$] F.Schr\"oeck, Quantum mechanics on phase space,
Dordrecht, Kluwer Academic Publishers (1996);\\ \indent
[15$^{\dagger}$] J.Perina, Z.Hradil, B.Jurco, Quantum optics and
fundamentals of physics, Dordrecht,
Kluwer Academic Publishers (1994);\\ \indent
[16$^{\dagger}$] About Heisenberg uncertainty relation, translated in
russian by A.Rogali, pp.210-217, in ``E.Schr\"odinger, Izbrannie
trudi po kvantovoi mehanike (Collected papers on quantum
mechanics)$^{\dagger}$, Moscow, Nauka (1976)";\\ \indent
[17$^{\dagger}$] G.Korn, T.Korn, Mathematical handbook for scientists
and engineers, sec. enl. and rev. ed., New York, N.Y., McGraw-Hill Publs.
Co., (1968) p.604, eq.(18.4-10);\\ \indent
[18$^{\dagger}$] H.Tucker, An introduction to probability and
mathematical statistics, New York, NY, Academic Press, (1963) p.91;\\ \indent
[19$^{\dagger}$] C.Gardiner, Handbook of stochastic methods, 2nd ed.,
Berlin, Springer-Verlag, (1985);\\ \indent
[20$^{\dagger}$] A.Angelow, Physica A, {\bf 256} (1998) pp.485-498,
the last equation in (24) and equation (B.4);\\ \indent
[21$^{\dagger}$] A.Angelow, Covariance, Squeezed and Coherent states: 
Proposal for experimental realization of Covariance states, Amer. Inst. Phys.,
Conference Proceedings BPU6 {\bf 899}, ISBN 978-0-7354-0404-5 (2007) pp.293-294;\\ \indent
[22$^{\dagger}$] A.Angelow, D.Trifonov, V.Angelov, H.Hristov, 
Method of Linear Invariants for description of beam dynamics 
of FEL undulator,  Proceedings of  9th Workshop, 
Nanoscience \& Nanotechnology,
Edited by E.Balabanova, I.Dragieva, Heron Press Ltd., Sofia, 2008 (accepted).\\
( available also in: http://arXiv.org/abs/0805.3626 )\\ \indent

\begin{thebibliography}{99}
\bibitem{a_star} $A^*$ is so defined that evidently
$A^*\Psi^*=(A\Psi)^*$, and therefore $A^*\Psi=(A\Psi^*)^*$.
\bibitem{slow} Of course, there are wave functions that, in a precise
given later moment, can determine the position with an arbitrary given
precision.
But one needs simply the chance to be able to trace back such an
approximation of a ``maximal function" by means of a wave equation over
the corresponding time interval, and the resulting function to be taken
as original state. But it should be not less than ``optimal".
\end{thebibliography}
\end{document}